# Improving Data Reusability in Interactive Information Retrieval: Insights from the Community

Tianji Jiang*
tianji008@ucla.edu
University of California, Los Angeles
Los Angeles, California, USA

Wenqi Li
Peking University
Beijing, China

Jiqun Liu
University of Oklahoma
Oklahoma City, Oklahoma, USA

## Abstract

In this study, we conducted semi-structured interviews with 21 IIR researchers to investigate their data reuse practices. This study aims to expand upon current findings by exploring IIR researchers' information obtaining behaviors regarding data reuse. We identified the information about shared data characteristics that IIR researchers needed when evaluating data reusability, as well as the sources they typically consulted to obtain this information. We consider this work to be an initial step towards revealing IIR researchers' data reuse practices and find out what the community need to do to promote data reuse. We hope that this study, as well as future research, will inspire more individuals to contribute to the ongoing efforts aimed at designing the standards, infrastructures, and policies, as well as fostering a sustainable culture for data sharing and reuse in this field.

## CCS Concepts

• **General and reference** → *General literature*; • **Information systems** → *Users and interactive retrieval*.

## Keywords

Interactive information retrieval, Research data management, Data reusability

## 1 Introduction

Information retrieval (IR) has a long tradition of reusing shared data sets and producing system-based experimental data for reuse by others [11, 13]. As IR research continues to evolve, the demand for large-scale high-quality data keeps growing [5]. Sharing and reusing research data can improve the cost-effectiveness of user studies and also facilitate reproducibility experiments, both are essential for the advance of IR research and evaluation techniques. In IR, whose research paradigm fundamentally rooted in comparison and evaluation, using common datasets also ensures the reliability, standardization, comparability, and reproducibility of study [5]. These benefits are not limited to system-centered evaluation. In Interactive IR, where the focus is on users' behaviors and tasks during their interactions with IR systems [7, 9], data sharing and reuse can be equally beneficial, just as in the broader IR domain.

The IIR research community has recognized the importance of sharing and reusing research resources, particularly research data, but also experiment designs and infrastructure. A recent empirical study found that data reuse is strongly advocated within the IIR community, although many practical challenges remain [?]. Another empirical analysis of CHIIR papers published between 2016 and 2022 showed that 14.4% of the papers reused existing data, indicating that data and resource reuse is no longer uncommon in IIR research [3]. Efforts to encourage and facilitate such practices have been an active topic within the community. The BIIRRR workshop series [1, 2, 4], as the most recognized efforts, examined barriers to resource reuse and identified key focus areas essential to advancing reuse. Building on these discussions, the same community distilled their insights into a manifesto published at CHIIR 2021, which highlights the principles, challenges, and requirements for documenting and archiving research to support future reuse [5]. In addition, there are also some explorations of practical approaches to facilitate or standardized resource sharing and reuse [6]. For example, Liu [8] developed Cranfield-inspired approaches to assess the reusability of IIR resources and to enhance reporting practices.

Building on our previous study of data reuse practices among IIR researchers [?], this study further examines a specific aspect of their data reuse experiences: *what characteristics of research data they consider important when deciding whether the data are reusable, and how they obtain information about these characteristics in practice*. Gaining a deeper understanding of these requirements and strategies will help clarify what "reusability" means for IIR research data and inform the design of knowledge infrastructures that better support the curation and sharing of such data. In addition, we anticipate that the findings of this study will contribute to ongoing efforts to foster a more seamless and effective ecosystem for circulating research resources within the IIR community.

## 2 Research Questions and Design

To contribute to the understanding of how data reusability is assessed in IIR research, this study addresses two key questions:

- **RQ1:** What characteristics of research data are required in the determination of their reusability?
- **RQ2:** How do researchers gather the information necessary to evaluate data reusability?

This study reused the interview data collected in our previous work [?], which involved in-depth, semi-structured interviews with 21 theoretically sampled IIR researchers. Participants varied in research interest but all had a minimum of three years of experience in IIR research to ensure a sufficient understanding of the domain and of what constitutes a rigorous study. The sample included eight professors, one postdoctoral researcher, and twelve senior Ph.D. students from geographically diverse institutions across Asia, Australia, Europe, and North America. Although the sample size is relatively small, the participants represent a broad range of research experiences, career stage, and institutional contexts. The interviews were conducted virtually. To ensure high-quality communication during the interviews, each interview was conducted in the participant's preferred language (either English or Chinese).

In the analysis phase, the authors employed thematic coding. All 21 interview transcripts were iteratively reviewed to identify segments relevant to the research questions. Conceptually similar segments were then grouped together and organized into thematic categories through constant comparison. These themes were subsequently refined and synthesized into key findings that directly address the research questions.

## 3 Findings

Across all participants, data reusability was not regarded as an inherent quality of the dataset itself but as a assessment constructed by the researcher who engages with the data. They repeatedly emphasize that decisions about data reuse were based on how they perceive and interpret the information available, from the dataset itself, documentation coming together with data, related literature, community norms, and their own expertise. Rather than viewing reusability as a fixed property, participants described it as a process of interpretation and evaluation, during which they gathered information about factors influencing a dataset's potential "fitness for reuse", based on the information available to them and their own knowledge (as demonstrated in Figure 1). As a participant described *"Once I get a (shared) dataset, I need to get more intimate and familiar with the collection based on my own experiences rather than just viewing it as a series of numbers." (P21)*.

**Table 1: Summary of findings**

| Data characteristics for evaluating reusability | Sources of information for data characteristics |
|---|---|
| • Context and methods for data production<br>• Documentation of data<br>• Creator credibility and community validation<br>• Legal and ethical constraints | • Academic literature related to the shared data<br>• Personal and professional networks<br>• Community venues and shared experiences |

In this paper, responding to the two research questions, the authors focus their analysis on the information that study participants sought when evaluating the reusability of shared data, and on the sources from which they obtained such information.

## 3.1 Characteristics IIR Researchers Consider When Evaluating Data Reusability

Participants considered various factors when determining whether, and to what extent, they could reuse datasets collected by other researchers for their own work. These factors included the context and methods for data production, the data documentation, data creators' credibility, and legal and ethical constraints.

*3.1.1 The context and methods for data production.* A nearly universal finding among the study participants (20 out of 21) was that understanding the context and methods of data collection is essential when assessing the reusability of shared datasets. Participants emphasized the need to know how, why, and under what conditions the data were originally collected. Such information is necessary for evaluating a dataset's trustworthiness, including whether it was produced through a rigorous and purpose-aligned research design, whether any issues in data collection may have introduced bias or other quality concerns, and whether the data can appropriately support their new research purposes. As P21 stated, *"If I don't know how it was collected and cleaned, I can't trust the data."*

This need for methodological clarity was especially strong among study participants who primarily work with datasets derived from lab studies examining how real users perform specific task types on given topics. For these researchers, behavioral data are contextually constructed, and the context of data production is inseparable from the data itself. This includes factors such as the experimental design, pre- and post-study interactions with participants, what users were asked to do during the study, and how they were instructed to perform those tasks. These elements are not ancillary details; they define the phenomenon under study. As another participant (P20) explained, *"User behavior depends on the specific interface. Without knowing the context, the data loses its meaning... The data was collected for their purpose. It cannot directly serve my research purpose if the task does not align."*

*3.1.2 The documentation of data.* Almost all participants emphasized that good documentation is indispensable for evaluating the reusability of shared data. Even when data were publicly available, many noted that they could not reuse them due to the lack of a clear description of what the dataset actually contains and what it was intended to demonstrate. When such information could not be easily inferred directly from the dataset itself, which often occurs with large-scale datasets or those containing many variables, the participants expected to obtain these details from the accompanying documentation. Participants described documentation as any form of accompanying material, including README files, metadata schema, data descriptors, or methods appendices, that provide interpretive guidance. Interestingly, participants offered varied focus on the content they expected to see in documentation, some of which are illustrated in Table 2.

Participants' expectations regarding what should be included in data documentation can be summarized as fulfilling three interrelated purposes. First, epistemic transparency, which refers to revealing how the data were produced and under what conditions. Second, pragmatic usability, which involves providing sufficient information for others to understand how to use the dataset so that analyses can be reproduced or the data effectively reused for a different purpose. Third, social trust signaling, which demonstrates the professionalism, competence, and accountability of both the data creators and the research process for which the data were originally collected. As the following subsection discusses, this last aspect constitutes another crucial concern in evaluating data reusability. Overall, data documentation of good quality, which was capable of fulfilling one or more of these purposes, was regarded by study participants as a clear symbol of a dataset's reusability.

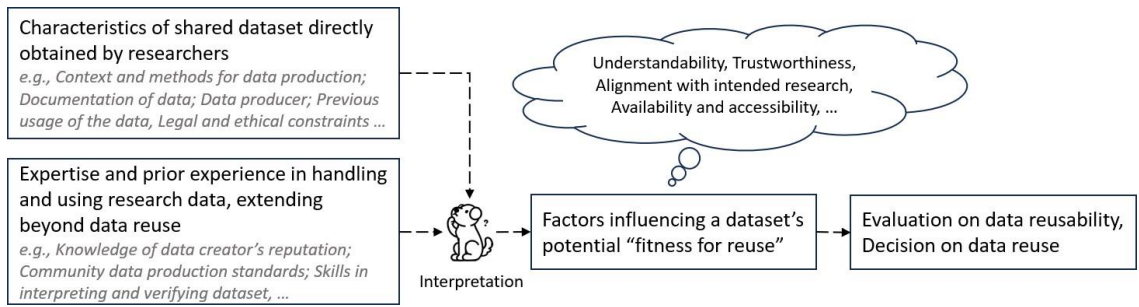

**Figure 1: The process of evaluating reusability of shared data**

**Table 2: Descriptions needed in the data documentation mentioned by study participants**

| Participant | Descriptions of data documentation |
| --- | --- |
| P1 | The documentation that explains the data structure is, at least for me, very important. In research, you need to know what information the data structure provides, what it does not provide, and how to use it appropriately. This is extremely important. |
| P4 | In its documentation, it often tells you how to use the data. Once you understand that, you can also form an overall sense of the dataset from a statistical perspective... If your research focuses more on short, everyday conversations, this kind of information helps you identify whether the dataset fits your needs. All of these impressions can be roughly obtained from its documentation. |
| P7 | In the data documentation, we want to precisely understand how the dataset was developed, what instructions were given, and what criteria the assessors used... such details are very important before reusing it. |
| P21 | Ideally we have README files. I'm a fan of just putting data in a CSV file ... but with a document that includes a link to the paper, how people should cite it, and an explanation of what each column is. |

*3.1.3 Creators' Credibility and Community Validation.* Another recurrent determinant of data reusability identified across participants was the credibility of the data creator and the extent to which the dataset had been validated through community use. Researchers rarely took a dataset "at face value". Instead, they made judgments of the dataset, mostly about trust concern, based on information not contained within the raw data or its accompanying documentation, such as who produced the dataset, who had reused dataset the venue in which it was shared, and how widely it had already circulated in their field.

It was found that trust in shared data often travels through reputation. Data originally collected by well-known researchers, reputable institutions, or individuals with strong research reputations, especially those known personally, were considered more reusable than data from less recognized sources. In contrast, datasets with unclear or unrecognized provenance tended to be viewed with suspicion, even when their documentation or related literature seemed to be technically sound. As described by P14: *"the authority and reputation of the original data collector could naturally give the impression that their data is of higher quality. A more reputable team might be seen as collecting data that is more accurate or reliable."*

The study participants experienced in evaluation-oriented IR study also described credibility as a kind of institutional guaranty. Data released by long-standing initiatives such as TREC or large corporate research groups were believed to be inherently more reliable and, of course, more reusable because they had passed through multiple layers of review, annotation, and reuse. As P19 noted, the strength of TREC collections lies not just in their design but in their repeated validation: *"They have gone through many rounds of iterations and a number of people have looked at them... we have the relevance judges, we have the queries — points of reference."* It was another noteworthy finding that participants who primarily worked with datasets derived from lab studies tied credibility to familiarity with the creator rather than to institutional guaranty. For them, trust in data was interpersonal. Because these researchers often followed different protocols in their own data collection and there was no community-wide standardized way to conduct such studies, they relied on their personal knowledge of the data creators to make judgments about methodological competence and ethical reliability. As a study participant (P11) noted: *"You have to know the team who did it. If I know them, I can trust their data; otherwise, I prefer to collect my own."*

*3.1.4 Legal and ethical constraints.* Several study participants also raised the issue of legal and ethical constraints as another key characteristic of shared data that concerned them, sometimes even determining whether reuse was possible at all. For many participants, particularly those experienced in working with human subjects or user behavior data, questions of consent, privacy, and data protection preceded any consideration of analytical potential. When examining a shared dataset, one of the first pieces of information they looked for and paid close attention to was the legal and ethical constraints surrounding its use. As one participant (P2) put it succinctly: *"The first thing I ask is: can I even ethically access this dataset, and how to appropriately make use of it? That's the gate before anything else."*

Furthermore, it was also mentioned that practices of ethical responsibility itself was a marker of research professionalism, which in turn contributed to the perceived reusability of shared data. Transparent handling of consent, access rights, and permitted uses not only protects the individuals from whom the data were originally collected but also signals the integrity of the data creators, thereby enhancing downstream trust and credibility. As mentioned by the study participants:

"*As long as the dataset clearly states what permissions and licenses apply, I feel confident in using it.*" (P17)

"*Good ethical documentation, like whether users consented or not and how to use it legally, shows that the creators took responsibility ... indicates the creator would be good researcher*" (P11)

## 3.2 The Source IIR Researchers Obtain Information to Evaluate Data Reusability

The absence of specialized repositories for behavioral or user-study data means that IIR researchers must piece together information about a shared dataset from a variety of sources to make informed judgments. As P8 described, "*There isn't really one place where you can browse user study datasets and information about them...*" In addition to examining the dataset itself and its accompanying documentation, participants mentioned several other sources of information, including academic literature related to the shared data, personal and professional networks, and community venues and shared experiences.

*3.2.1 Academic literature related to the shared data.* Most participants (n=20) reported that academic publications were the main, and often the only, source of information available to them about existing datasets. This source was equally accessible to anyone reviewing their own studies that reused these data, which in turn allowed participants to feel more confident about the transparency of their work, since they could evaluate the data based on the same information available to reviewers. Participants also noted that original data creators often included the most detailed accounts of their data production processes in academic papers, not necessarily to facilitate reuse but to ensure transparency in reporting their study design. By consulting the methods sections, appendices, and supplementary materials, participants inferred how the data were collected, what tasks participants performed, and whether the study design aligned with their own research purposes. As P11 described, *"Mainly by reading papers ... There's no other channel ... I read how they designed the task, what kind of participants they had, and then decide if it fits."*

*3.2.2 Personal and professional networks.* Personal networks and collaborations play a crucial role in learning about data characteristics to evaluate reusability. Researchers often rely on colleagues, supervisors, or conference connections to understand shared datasets or to confirm their credibility. As P15 described her experience of obtaining information of shared dataset:*"It's often through people. If I know the lab or the author, I feel confident using their data ... I could contact the data creator when I was not sure about something."*

*3.2.3 Community venues and shared experiences.* Community venues such as CHIIR and SIGIR also served as indirect mechanisms for discovering and obtaining information about others' shared data. In particular, the establishment of specialized tracks for resource sharing at these conferences has encouraged researchers not only to share their datasets and experiences but also to learn where to find shared data and related insights. Through conference presentations and informal discussions, researchers often learned about datasets produced by others and gained first-hand understanding of their contents and contexts. As P20 described: "*Usually at conferences, someone presents a dataset. If it's interesting, I might contact them later to see if it's available and ask (further) questions I still had regarding the dataset.*"

These sources of community venues and shared experience, published literature, and professional networks, show that in IIR, learning about and evaluating shared datasets is closely tied to community interactions rather than to formal repositories. Knowledge about datasets circulates through conference presentations, informal conversations, and academic publications, where researchers exchange not only data but also the contextual understanding needed to interpret them. In this way, the community itself functions as a living repository, a network of people, papers, and practices through which data-related knowledge is continually produced, validated, and transmitted. Recognizing this pattern highlights the need to design knowledge infrastructures that build on existing community dynamics by making these informal exchanges more visible, persistent, and accessible, thereby supporting responsible data sharing and reuse within the IIR field.

## 4 Discussion and Conclusion

As in all scientific endeavors, progress in IIR research is contingent on the ability to build on previous ideas, approaches, and resources [1, 10]. Practices and discussions have been made to establish shared disciplinary repositories of research tools and data, summarize and analyze the current data reuse practices, as well as promote data sharing and reuse in IIR. The research community have identified a number of barriers to reuse of resources in IIR research, including the fragmentary nature of how the community's resources are organized, the lack of awareness of their existence, and insufficient documentation and organization of the resources available [2]. However, the current discussions on the approaches and barriers to data reuse in IIR research are primarily idea-generating, while the field needs more in-field explorations to provide evidence and further insights to support these discussions. This study aims to expand upon current findings by exploring IIR researchers' data and resources seeking behaviors for supporting data reuse. We found the information about certain characteristics of shared data the IIR researchers considered when evaluating data reusability, and the source through which the study participants usually obtained such information. Our study contributes to ongoing efforts to understand the practices, challenges, and decision-making processes surrounding data sharing and reuse within the IIR community. Furthermore, the results point to opportunities for developing knowledge infrastructures that build upon existing community practices (e.g. TREC and similar venues [e.g. 12]), making data-related knowledge more visible, connected, and accessible, and thereby supporting more effective and responsible data sharing and reuse in this field.

This study has potential limitations. First, the findings are based on participants' self-reported thoughts and behaviors regarding data reuse. We recognize the possibility that these reports may differ from the participants' actual practices in their research. Second, as the study adopts a semi-structured approach, the conversations with participants may be influenced by the interviewer's interests, which could introduce biases and affect our findings. In future studies, we plan to conduct on-site observations of the' data discovery and reuse behaviors of IIR researchers to gain a better understanding of their expectations, concerns, and reuse practices.